# FABRICATION OF FLEXIBLE SUPER CAPACITOR USING LASER LIGHTSCRIBE TECHNIQUE

Gladson Joseph[1], A.Alfred Kirubaraj[2], U.Satheesh[1], D. Devaprakasam[1]
NEMS/MEMS/NANOLITHOGRAPHY Lab
[1]Department of Nanosciences and Technology,
[2]Department of Electronics and Communication Engineering,
Karunya University, Coimbatore 641114, India.
Email:devaprakasam@karunya.edu

**Abstract:** Super-capacitors are promising energy storage devices due to their capability of delivering high peak current and storing huge amount of energy in a short time with very low internal power loss. We fabricated the graphene/graphite oxide super-capacitor using laser Lightscribe technique. We prepared graphite oxide by modified hummers method and used PET film as a flexible substrate on which graphite oxide (GO) was coated. Using Lightscribe drive and software, the super-capacitor configuration was patterned on the GO coated PET film. During the writing process, the laser converts GO into graphene. We characterized the fabricated flexible super-capacitor which exhibits high resistance of 20KΩ with applied voltage of 10V and further increase of voltage (20V) decreases the resistance to 8KΩ. We also analysed the frequency response of the capacitor using impedance measurement which shows high frequency response and estimated capacitance is 120nF. We optimized the patterns by running the Lightscribe repeatedly on the GO coated PET substrate.

**Keywords:** Super-capacitor, Graphite Oxide (GO), polyethylene terephthalate (PET), Impedance Spectroscopy, Cyclic Voltametry

## INTRODUCTION

A super-capacitor is a specially designed capacitor having large capacitance and is energy storing devices delivering high peak currents. A capacitor consists of two parts, conducting and insulating regions. The conducting region will act as an electrode and the insulating region will be the dielectric media and the device produced with a capability of storing relatively higher energy in the form of electrical charge. This kind of device has very short charging and discharging time called as super-capacitor. Energy density ratio for super-capacitor typically ranges from 0.5 and 10 W-h/kg which is considerably higher than that of a standard capacitor. The major problem of AC super-capacitor is that it lacks the energy density [1, 2]. Figure 1 shows the active carbon materials are the best suitable for capacitor electrodes. The theoretical studies report that the capacitance of the electrical double layer capacitor is about 550F/g [3].

Considering the superior electrical, mechanical and thermal properties of grapheme and graphite oxide are the promising materials and they are used in energy storage applications as electrode and dielectric material. The conductivity of graphite oxide is between 1 and $5\times10^{-3}$ S/cm. We can modify the property of the graphite, GO and graphene by oxidation and reduction process [4].

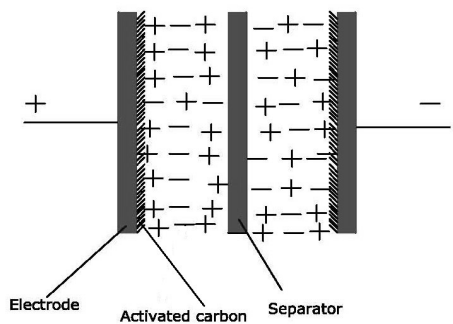

Figure 1 shows the electrochemical double layer capacitor.

When the super-capacitor is powered, it creates an electron cloud near one electrode and an electron deficient region near another electrode which will help to store charge in a capacitor. The energy storage in a super capacitor depends on non-faradic charge. In this process there will not be any charge transfer across the electrode.

In this work, (i) in section 2 deals about Light scribe Technology, (ii) in section 3 deals about fabrication and experimental techniques of super capacitor, (iii) in section 4 deals about result and discussion, and (iv) Section 5on conclusion.

## LIGHTSCRIBE TECHNOLOGY ®

Lightscribe technology was invented by Hewlett-Packard for direct disc labeling process. In this process, the surface color of the disc will change to a darker shade when exposed to light. The wavelength of the optical system uses 788nm which will create etches on the surface of the Lightscribe disc. A Lightscribe enabled DVD drive uses an optical laser to burn an image into the thin dye coating on the label side of a Lightscribe disc. The Lightscribe labeling system has no ink to smear, no paper to curl, and no sticky adhesive to cause problems.

### FABRICATION AND EXPERIMENTAL TECHNIQUES OF SUPER-CAPACITOR

### Synthesis of Graphite Oxide

The synthesis of graphite oxide was done by modified Hummers method. 2gm of pristine graphite powder was added to 50ml of concentrated $H_2SO_4$ and stirred for 30 minutes at room temperature. Then 6gm of $KMnO_4$ powder was added slowly to the above mixture and stirred for 8hrs to avoid the heating problems.5ml of $H_2O_2$ was added to the above mixture immediately followed by the addition of 90ml of water and further stirred for 3hrs.300ml of water was added to the solution and kept for sedimentation for 1 day. Except the sediment water, the remaining content was poured out. 5% of HCl was added and stirred for 2hrs and the solution was allowed to settle for 2 days. After settling, solution was centrifuged for about 15 times until its pH valuereaches 7 (neutral). The extracted sediment was dried at 60˚C for 24hrs to get a pure graphite oxide powder [6].

### Super Capacitor Device Fabrication

Poly ethylene terephthalate (PET) film is used as the base substrate for the super capacitor device and is then washed with acetone to make the surface clean. GO solution is coated above the surface of the PET film and dried at ambient conditionsto form a thin layer of graphite oxide over the film. The coated PET film is mounted on the surface of the Lightscribe DVD disc. Then the drawn super-capacitor pattern was written on the coated PET film by Lightscribe laser.

## EXPERIMENTAL RESULTS AND DISCUSSION

### SEM analysis of graphite oxide

Figure 2 shows the SEM image of graphite oxide. The flake like structure of GO is clearly observed, the flake dimension is about 1µm.

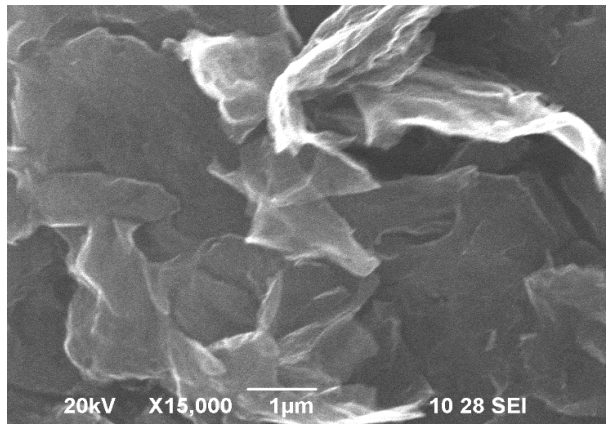

Figure 2 shows the SEM image of graphene oxide layer.

### XRD characterization

Figure 3 shows the XRD characterization of graphite oxide. From the XRD pattern, we observe the high intensity peak exist at 10 degree confirming that the prepared material is graphite oxide.

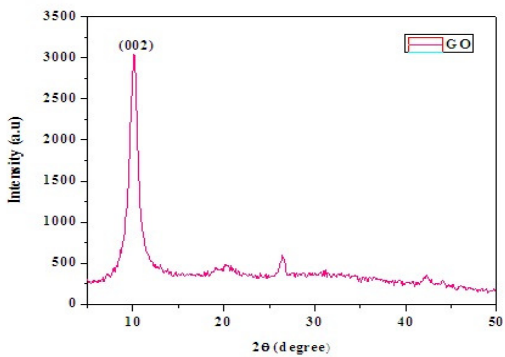

Figure 3 shows the XRD characterization of graphite oxide.

### Optical image analysis of super capacitor

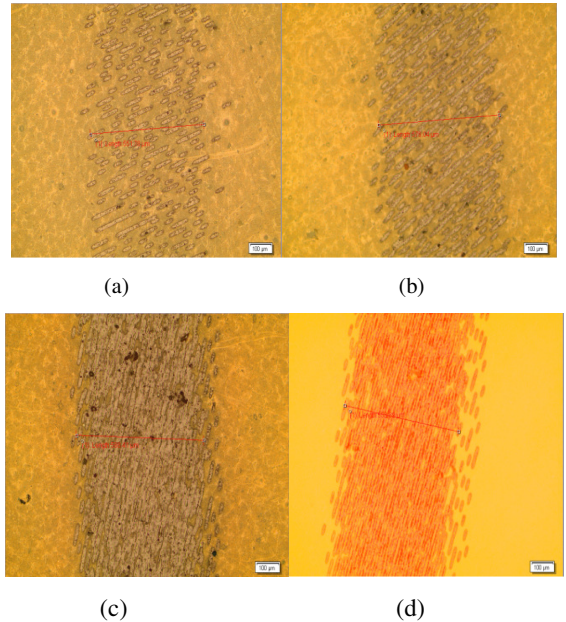

(a)  (b)

(c)  (d)

Figure 4(a-d) shows the optical image of super-capacitor for different Lightscribe cycle.

From the optical microscopy, we measured the scribed patterns for the Lightscribe cycle of 1, 3, 6 and 8.After the 8[th] cycle the scribed track become continuous and the final width is about 550micron. From this optimization process, the conversion of graphene oxide got improved on repeated labelling process.

### RAMAN spectroscopic characterization

We investigated the prepared graphite oxide using Raman spectroscopy and observed both D and G vibrations, the G vibration occurred at 1593 cm$^{-1}$ and D vibration observed at 1355 cm$^{-1}$(Figure 5).

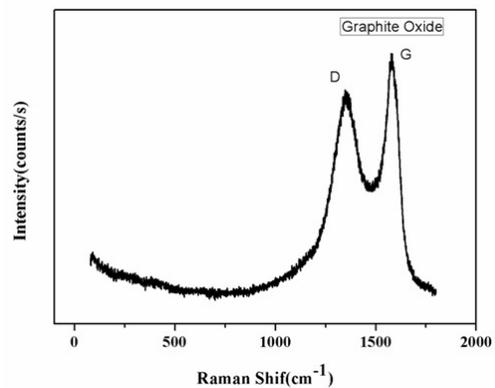

Figure 5 shows the Raman spectra of graphite oxide.

After the laser light scribing, the converted graphene was investigated, the Raman spectrum (Figure 6) clearly shows that the graphite oxide become graphene.

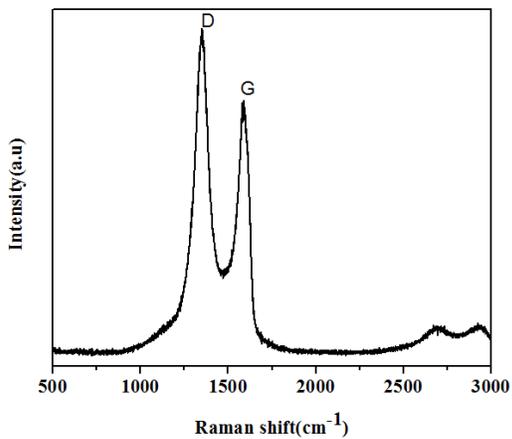

Figure 6 shows the Raman spectra of graphene.

From the Raman spectra of the graphene (Figure 6), we observed two main peaks in Raman spectrum caused by laser excitation. G mode of vibration observed at 1580 cm$^{-1}$, a plane vibration mode 2D at 2690 cm$^{-1}$, and second order overtone of a different inclined vibration D at 1350 cm$^{-1}$. D and 2D peak position are dispersive.

**I-V Characterization of super capacitor**

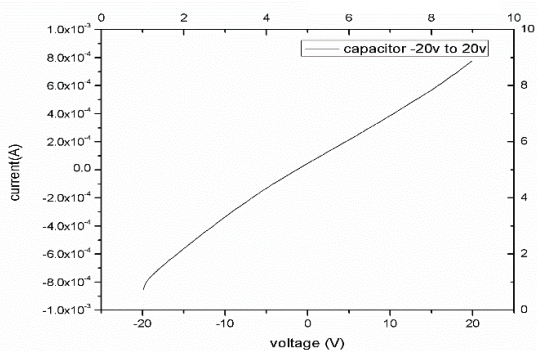

Figure 7 shows the I-V characterization of graphite oxide capacitor.

From the above characterization (Figure 7), we obtain the following nature of flexible super capacitor as (i) at low voltage, the resistance of the device is more and (ii) at high voltage, the resistance of the device is less.

**Impedance Characteristic**

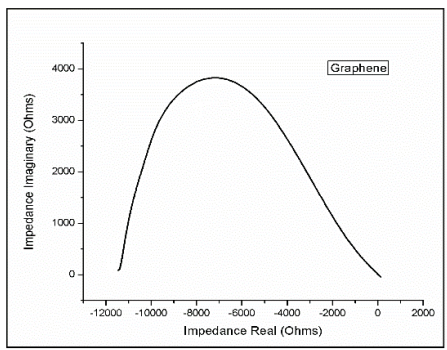

(a)

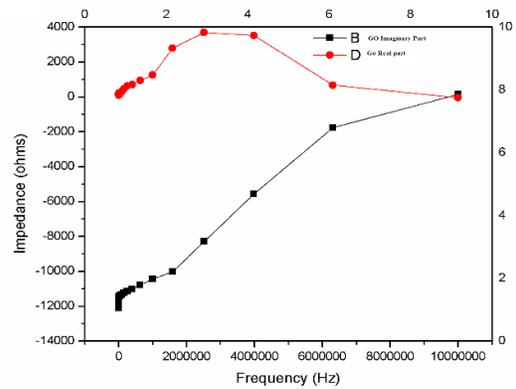

(b)

Figure 8 (a, b) shows the impedance characteristic of super-capacitor, real impedance response to the frequency.

Figure 8(a) shows the real versus imaginary impedance for the varied voltage (0.01mV to 1mV) and frequency ratio, which works at a frequency range from 1Hz to 10 KHz. At the frequency, the estimated resistance is about 3.8MΩ. Figure 8(b) shows the frequency response of the designed super-capacitor which operates at very high frequency 10GHz. This kind of capacitor is very useful in high frequency circuits.

**Fabrication of Super Capacitor in Lightscribe Disc**

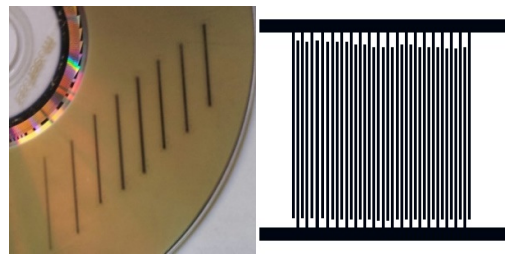

(a)                          (b)

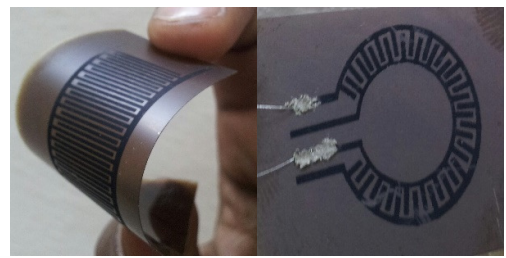

(c)                          (d)

Figure 9 shows (a) Lightscribe patterns at different cycles (b) Image of the pattern, (c) final structure of flexible super-capacitor, (d) circular design of flexible super-capacitor.

In Figure 9(a), image of a labelled disc is shown; in which the depth of each line is varied. The reason for this variation is due to the difference in number of labelling cycles. If we scribe the same pattern for a repeated times, the depth of the etched pattern will be more and the formation of the pattern is continuous. The first line has undergone a single scribing process and the depth of the pattern is very low. To get a continuous pattern over the substrate, the labelling cycle is repeated for more than 8 times. Figure 9(b)shows the image of the device pattern drawn with the help of Lightscribe software tools. Figure9(c, d) are the images of the flexible super-capacitor.

## CONCLUSION

The flexible GO coated super-capacitor is fabricated using Lightscribe technology. We analysed the different properties of the GO coated super-capacitor device which is similar to the super-capacitor. The capacitor shows a capacitance of about 120nF and the resistance of the device is low (8kΩ) in high frequency (8 KHz to 10GHz). This kind of capacitor is very useful in high frequency circuits. In future, we can improve the performance of the device by reducing the wavelength of the Lightscribe laser to get a fine resolution of patterns. Our study demonstrates that using Lightscribe technology flexible devices and sensors can be fabricated and the complex nano structure can be implemented.

## ACKNOWLEDGEMENT

We thank DST-Nanomission, Government of India and Karunya University for providing the financial support to carry out the research. We also thank the department of Nanosciences and technology for the help and support to this research.